\documentclass{article}

\usepackage{arxiv}

\usepackage[utf8]{inputenc} 
\usepackage[T1]{fontenc}    
\usepackage{hyperref}       
\usepackage{url}            
\usepackage{booktabs}       
\usepackage{amsfonts}       
\usepackage{nicefrac}       
\usepackage{microtype}      
\usepackage{lipsum}
\usepackage{caption}
\usepackage{graphicx}
\usepackage{subcaption}
\usepackage{enumitem}
\title{ABOME: A Multi-platform Data Repository of Artificially Boosted \\Online Media Entities}
\newcommand{\dbname}{\texttt{ABOME}}

\author{
    Hridoy Sankar Dutta \\
  IIIT Delhi, India \\
  \texttt{hridoyd@iiitd.ac.in} \\
   \And
 Udit Arora\\
 IIIT Delhi, India \\
  \texttt{udita@iiitd.ac.in} \\
   \AND
   Tanmoy Chakraborty \\
   IIIT Delhi, India \\
   \texttt{tanmoy@iiitd.ac.in } \\
}

\begin{document}
\maketitle

\begin{abstract}
The rise of online media has incentivized users to adopt various unethical and artificial ways of gaining social growth to boost their credibility within a short time period. In this paper, we introduce \dbname, a novel multi-platform data repository consisting of artificially boosted (also known as blackmarket-driven {\em collusive entities}) online media entities such as Twitter tweets/users and YouTube videos/channels, which are prevalent but often unnoticed in online media. \dbname~allows quick querying of collusive entities across platforms. These include details of collusive entities involved in blackmarket services to gain artificially boosted appraisals in the form of {\em likes}, {\em retweets}, {\em views}, {\em comments}, {\em follows} and {\em subscriptions}. \dbname~contains data related to tweets and users on Twitter, YouTube videos and YouTube channels. We believe that \dbname~is a unique data repository that can be used as a benchmark to identify and analyze blackmarket-driven fraudulent activities in online media. We also develop \texttt{SearchBM}, an API and a web portal that offers a free service to identify blackmarket entities.
\end{abstract}


\section{Introduction}
The past decade has seen a momentous rise in Online Social Networks (OSNs) such as Twitter, YouTube, and Facebook, which help people connect for personal and business interactions. These platforms now boast billions of active users, thereby making them an important component of today's human social fabric. People share and form thoughts and opinions about events, products, and other people on these platforms. This makes online media an attractive platform for people who wish to propagate their opinions to spread their agenda, such as promoting a product or political ideology. Therefore, gaining a stronger influence on online media platforms carries a high level of economic benefit.

However, in order to spread an opinion, users require a large reach across the network. This reach can either be acquired {\em organically} -- by posting quality content over time and gaining popularity, or {\em inorganically} -- by certain online media-driven blackmarket services that allow users to boost the reach of their content artificially. \textit{Collusion in online media} involves users artificially gaining social reputation, which violates the Terms of Service (ToS) of the online media platform. These users approach blackmarket services to artificially inflate their social status. This results in entities such as Twitter tweets/users or YouTube videos/channels to appear more attractive to the end-users, thus leading to activities such as fake promotions, campaigns, and misinformation. The blackmarket services support various online media services ranging from online social networks to other platforms such as rating/review platforms, video-sharing platforms and even recruitment platforms \cite{dutta2020blackmarketsurvey}.

A substantial number of studies have investigated the phenomena of {\bf fake}~\cite{alsaleh2014tsd,gupta2013faking,gupta2019malreg,cresci2015fame,stringhini2013follow}, {\bf fraudulent}~\cite{giatsoglou2015retweeting,liu2017holoscope,li2016world,shah2014spotting,hooi2016birdnest,chavoshi2016debot} and {\bf spam}~\cite{benevenuto2010detecting,yardi2010detecting,thomas2011suspended} activities. We encourage the readers to go through  \cite{kumar2018false,pierri2019false,da2019can} for detailed surveys on false and fake information on the web.

However, there are relatively fewer studies on the detection and analysis of collusive activities that result in an artificial boosting of social growth. Our recent investigations \cite{chetan2019corerank,dutta2018retweet,dutta2019blackmarket,arora2019multitask,dhawan2019spotting,dutta2020hawkeseye,arora2020analyzing,sankar2020detecting} revealed that existing fraud detection strategies are not suitable for blackmarket-driven collusive entity detection. These studies reported that collusive users are not bots or fake users; rather, they are normal users showing a mix of organic and inorganic activities with no synchronicity across their behaviors. \dbname~would help researchers analyze blackmarket-driven collusive activities and build systems to detect them. 

\dbname~consists of multi-platform datasets for collusion in online media collected from two major blackmarket services - YouLikeHits and Like4Like. \dbname\ comprises datasets of two types -- \textit{historical data} and \textit{time-series data}. The former type of dataset is divided into two parts -- the first part consists of 23,522 collusive retweets and 18,368 collusive follower requests for Twitter; the second part consists of 58,091 collusive likes, 25,106 comments, and 7,847 subscriptions requests for YouTube. The latter type of dataset consists of time-series data of 2,350 Twitter users and 4,989 tweets collected from blackmarket services.

\dbname~is unique for the following five reasons:
\begin{itemize}
\item To the best of our knowledge, \dbname~is the first public dataset of collusive entities in online media such as Twitter tweets/users and YouTube videos/channels. We believe these datasets have tremendous research potential in the field of analysis and detection of collusive behavior in online media.
\item \dbname~comprises of two types of datasets: \textit{historical} and \textit{time-series} data.
\item \dbname~provides abundant textual and temporal information of collusive entities for Twitter and YouTube.
\item \dbname~has an API and a web portal, \texttt{SearchBM} to discover collusive entities using text search queries.
\item \dbname~protects user privacy and can be used in a wide range of research areas, such as fraudulent entity detection, diffusion modeling, social-growth prediction, etc.
\end{itemize}
\noindent\begin{minipage}[t]{.95\columnwidth}
The entire dataset, along with a smaller sample, is available at the following link: \href{https://zenodo.org/record/4437987}{https://zenodo.org/record/4437987} \\(Dataset DOI: http://doi.org/10.5281/zenodo.4437987). We also provide a datasheet for our dataset according to Datasheets for Datasets recommendations \cite{gebru2018datasheets} as supplementary material.
\end{minipage}

\section{Blackmarket Services} 
Websites such as YouLikeHits (\url{https://www.youlikehits.com/}), Like4Like (\url{https://www.like4like.org/}), TraffUp (\url{https://traffup.net/}), JustRetweet (\url{https://www.justretweet.com/}) allow social media users to gain appraisals  {\em inorganically} in different forms. They provide services for various \textbf{online social networks}, e.g., Facebook (followers, likes, shares, comments), Twitter (followers, retweets, likes), Instagram (followers, likes, comments). Other than OSNs, the blackmarket syndicates also provide service to \textbf{video subscription-sharing platforms}, e.g., YouTube (views, subscribers, likes, comments), Vimeo (plays, followers), \textbf{music-sharing platforms}, e.g., SoundCloud (plays, followers, likes, reposts, comments), ReverbNation (fans), \textbf{business and employment-oriented platforms}, e.g., LinkedIn (followers, connections, endorsements).
\begin{figure}[!htbp]
    \centering
    \includegraphics[width=\columnwidth]{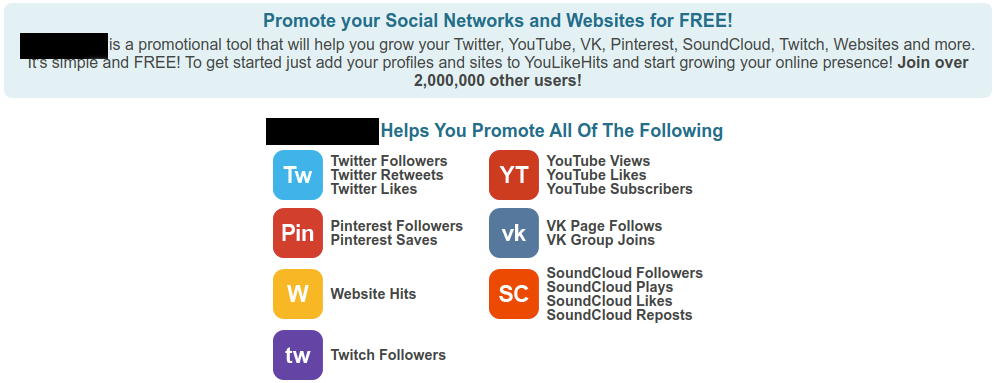}
    \caption{Example blackmarket service providing collusive appraisals to online media platforms such as Twitter, Pinterest, YouTube, VK, SoundCloud, Twitch (name of the blackmarket redacted). }
    \label{fig:example_bm}

\end{figure}
Organically gaining higher reach on online media is difficult and time-consuming, which boosts the allure of these blackmarket services. With higher reach comes stronger influence, and with stronger influence comes higher economic value. This influence can be used for product promotions or opinion propagation. Fig. \ref{fig:example_bm} shows an example of one such blackmarket service which provides collusive appraisals.

\cite{shah2017many} divided the blackmarket services into two types based on the model of service - \textit{Premium} and \textit{Freemium}. Customers in the premium services
need to pay a cash lump sum to receive blackmarket services.  Freemium services may not demand customers to pay; rather, these services create a community of customers where each member gains appraisals by appraising the content of other customers registered on these blackmarket services. When a customer appraises some content through the blackmarket portal, they earn \textit{credits}, which they can use later to gain appraisal for their own content. Such freemium services are also called {\em credit-based freemium services}. These services are a menace and pose a threat to the credibility of social media platforms. Unlike bots, detecting users who engage in these activities is difficult because they may display a mix of organic and inorganic behavior - whereby they appraise some content related to their interest as a genuine user and also appraise some content to gain credits on a freemium service.

We  made the first attempt to investigate blackmarket customers on Twitter \cite{dutta2018retweet}. We collected data related to users engaged in producing fake retweets from four blackmarket services and annotated each user into one of the four categories -- \textit{bots, promotional customers, normal customers, and genuine users}. Finally, we ran several classifiers on a set of 64 features to perform multi-class and binary classification to detect collusive users. We further extended this work to show how users involved in premium blackmarket services exhibit unusual properties as compared to those involved in freemium services \cite{dutta2019blackmarket}. \cite{chetan2019corerank} proposed \texttt{CoReRank}, an unsupervised approach to detect collusive users and suspicious tweets submitted to blackmarket services based on two intrinsic traits -- the credibility of users and the merit of tweets. \texttt{CoReRank} considers a directed bipartite graph of (re)tweeters and (re)tweets in order to incorporate interdependency of user-level and content-level traits such as network properties, behavioral properties, and topical similarity. \cite{arora2019multitask} detected tweets submitted to blackmarket services using a multitask learning approach. \cite{arora2020analyzing} proposed a multi-view learning-based approach to detect collusive retweeters by utilizing various attribute and network views based on the user's posts and interactions on the social graph. We encourage the reader to go through \cite{dutta2020blackmarketsurvey} for a comprehensive survey on analyzing and detecting collusive activities in online media platforms.
\begin{figure*}[!htbp]
    \centering
    \includegraphics[width=\textwidth]{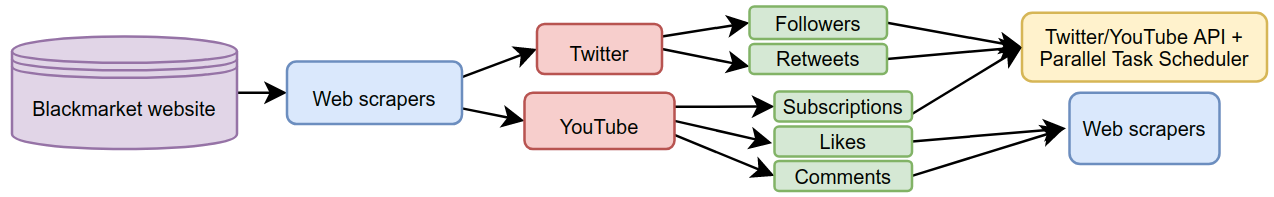}
    \caption{The process of collecting \dbname~dataset from the blackmarket service.}
    \label{fig:data_collection}
\end{figure*}
\section{Data Collection}
In this section, we first comment on our ethics and data privacy statement. We then describe the process of creating our datasets in detail.
\subsection{Ethics and Data Privacy Statement}
The entire data collection process has been carried out through Twitter API\footnote{\url{https://developer.twitter.com/en/docs}}, YouTube API\footnote{\url{https://developers.google.com/youtube/v3}} and web scrapers. We did not seek explicit permission from YouLikeHits and Like4Like to scrape the content because these blackmarket sites do not themselves act in line with the Terms \& Conditions (T\&C) of the services they connect with\footnote{We don't reveal the identity of users/tweets (Twitter id, Tweet ids, YouTube channel ids, etc.)}. They allow users to boost their influence on these platforms artificially, thereby going against the T\&C of Twitter\footnote{\url{https://help.twitter.com/en/rules-and-policies/platform-manipulation}} and YouTube\footnote{\url{https://support.google.com/youtube/answer/3399767?hl=en}}. Further, since the data we posted is anonymized, the privacy and identity of these users will not be compromised. We abode by the terms, conditions, and privacy policies of our Institute Institutional Review Board (IRB) approval\footnote{Permission to collect data from blackmarket services is mentioned in our IRB approval.}.

\subsection{Collecting Data from Blackmarket Services}
After careful IRB approval, we developed web scrapers for parsing two of the most popular blackmarket websites - YouLikeHits and Like4Like. The parser for YouLikeHits used Python's BeautifulSoup library to parse the HTML DOM of the website and got the details of the users and content that are posted for appraisals. The parser for Like4Like used Selenium (\url{https://www.seleniumhq.org/}) to run a headless web browser within which the website is loaded and BeautifulSoup was then used to parse the details of users and content posted for appraisal.


\subsection{Data Anonymization}
The data is anonymized by removing all Personally Identifiable Information (PII) and generating pseudo-IDs corresponding to the original IDs. A consistent mapping between the original and pseudo-IDs is used to maintain the integrity and usefulness of the data.

\subsection{Collecting Data at Scale from Twitter and YouTube}
We focused on collecting data from credit-based freemium services. We divide the datasets into two parts:

\begin{itemize}
    \item \textbf{Historical data}:
    This consists of all the data for Twitter and YouTube from YouLikeHits gathered via sequential querying of the website's URLs (the historical data for Like4Like was not available, which is why it is missing in our collected dataset) between the period March-June, 2019. The details of the sequential querying technique are explained in the last part of this section.
    \item \textbf{Time-series data:}
    This consists of time-series data (collected every 8 hours) of Twitter users and tweets collected from two blackmarket services -- YouLikeHits and Like4Like between the period of March-June, 2019.
\end{itemize}

\subsubsection*{Historical data:}
Since the entities we collected were a few years old in many cases, we focused on collecting the relatively static properties (such as the profile description and tweet content) and information related to those entities. We collected the following historical data from YouLikeHits:

\begin{enumerate}
    \item \textbf{Twitter Retweets:} Tweet ids of tweets that have been posted on YouLikeHits in order to gain retweets, and their tweet objects using the Twitter API as well as the last 100 retweets of these tweets (we were only able to collect the last 100 retweets due to Twitter API limitations).
    \item \textbf{Twitter Followers:} User ids of accounts that have been posted on YouLikeHits in order to gain followers, and their user objects using the Twitter API.
    \item \textbf{YouTube Likes and Comments:}  Video ids of videos that have been posted on YouLikeHits in order to gain likes, and their corresponding metadata, which is detailed in the next section.
    \item \textbf{YouTube Subscriptions:} Channel ids of YouTube channels that have been posted on YouLikeHits in order to gain subscribers, and their corresponding metadata (which is detailed in the next section).
\end{enumerate}

\subsubsection*{Time-series data:}
Since the entities we collected here were recent and fetched periodically, we collected dynamic information (such as the follower/followee network of Twitter users) related to these entities along with the static information. To collect time-series data, we developed a parallel task scheduler that can use multiple Twitter API keys to fetch a large volume of data at high speed. The task scheduler used the Tweepy library in Python for the API requests and ran in parallel the requests being made using the Python multiprocessing module. Multiple processes were created, and each process was assigned a given API key to work with. The code for the Parallel Task Scheduler is available at: \url{https://tinyurl.com/y2ztmoqg}. We used the Parallel Task Scheduler to collect time-series data after every 8 hours. We collected the following data between the period of March -- June 2019: 

\begin{enumerate}
    \item \textbf{Retweets:} We parsed the tweet ids of the tweets that have been posted by blackmarket customers in order to gain retweets of their own tweets, and collected the tweet objects, retweets of these tweets, and the timelines of the authors of these tweets.
    \item \textbf{Twitter Followers:} We parsed the user ids of the user accounts which have been posted by blackmarket customers in order to gain followers on their accounts and collected their timelines, follower and followee networks.
\end{enumerate}

The scheduler was designed in such a way that it was able to collect data for all Twitter entities after every $k$ hours. In our case, we used $k = 8$ to collect data every 8 hours and ignored users who had more than 50,000 followers or followees due to the Twitter API rate limits. Figure \ref{fig:data_collection} summarizes the steps followed to produce \dbname~dataset from the blackmarket services.

\section{Data Description}
We release two different types of datasets as a part of \dbname~- historical data and time-series data, as explained in the previous section.

\subsection{Historical Data}
We collected the metadata of each entity present in the historical data.
\paragraph{{\bf Twitter:}} We collected the following fields for retweets and followers on Twitter:

\textbullet{} \texttt{user\_details}: A JSON object\footnote{\url{https://developer.twitter.com/en/docs/tweets/data-dictionary/overview/user-object}} representing a Twitter user.

\textbullet{} \texttt{tweet\_details}: A JSON object\footnote{\url{https://developer.twitter.com/en/docs/tweets/data-dictionary/overview/tweet-object}} representing a tweet.

\textbullet{} \texttt{tweet\_retweets}: A JSON list of tweet objects representing the most recent 100 retweets of a given tweet.
\begin{table}[!htbp]
\centering
\begin{tabular}{ |c|c|c|c| } 
 \hline
 \textbf{User Type} & \textbf{Total} & \textbf{Suspended} & \textbf{Verified} \\ 
 \hline
 Retweet requests & 36,029 & 12,507 & - \\ 
 \hline
 Follower requests & 23,152 & 4,784 & 114 \\
 \hline
\end{tabular}
\caption{Summary of Twitter users for which historical information was collected from Freemium blackmarket services.}
\label{table:historical_twitter}
\end{table}

The details of the fields obtained from Twitter API can be found in the Twitter API Documentation (\url{https://developer.twitter.com/en/docs.html}).

\paragraph{{\bf YouTube:}} We collected the following fields for YouTube likes and comments:



\textbullet{} \texttt{is\_family\_friendly:} Whether the video is marked as family friendly or not.

\textbullet{} \texttt{genre:} Genre of the video.

\textbullet{} \texttt{duration:}  Duration of the video in ISO 8601 format (duration type). This format is generally used when the duration denotes the amount of intervening time in a time interval.

\textbullet{} \texttt{description:} Description of the video.


\textbullet{} \texttt{upload\_date:} Date that the video was uploaded.

\textbullet{} \texttt{is\_paid:}  Whether the video is paid or not.


\textbullet{} \texttt{is\_unlisted:} The privacy status of the video, i.e., whether the video is unlisted or not. Here, the flag \textit{unlisted} indicates that the video can only be accessed by people who have a direct link to it.

\textbullet{} \texttt{statistics:} A JSON object containing the number of dislikes, views and likes for the video.

\textbullet{} \texttt{comments:} A list of comments for the video. Each element in the list is a JSON object of the text (\textit{the comment text}) and time (\textit{the time when the comment was posted}). 
\begin{table}[!htbp]
\centering
\begin{tabular}{ |c|c|c| } 
 \hline
  \textbf{Type} & \textbf{Total} & \textbf{Suspended}\\ 
 \hline
 Like requests & $69200$ & $11109$\\ 
 \hline
 Comment requests & $30131$ & $5025$ \\
 \hline 
 Subscription requests & $11282$ & $3435$ \\
 \hline
\end{tabular}
\caption{Summary of YouTube videos and channels for which information was collected from Freemium blackmarket services.}
\label{table:freemium_youtube}
\end{table}

 We collected the following fields for YouTube channels:

\textbullet{} \texttt{channel\_description:} Description of the channel.

\textbullet{} \texttt{hidden\_subscriber\_count:} Total number of hidden subscribers of the channel.

\textbullet{} \texttt{published\_at:} Time when the channel was created. The time is specified in ISO 8601 format (YYYY-MM-DDThh:mm:ss.sZ).

\textbullet{} \texttt{video\_count:} Total number of videos uploaded to the channel.

\textbullet{} \texttt{subscriber\_count:} Total number of subscribers of the channel.

\textbullet{} \texttt{view\_count:} The number of times the channel has been viewed.

\textbullet{} \texttt{kind:} The API resource type (e.g., \textit{youtube\#channel} for YouTube channels).


\textbullet{} \texttt{country:} The country the channel is associated with.


\textbullet{} \texttt{comment\_count:} Total number of comments the channel has received.

\textbullet{} \texttt{etag:} The ETag of the channel which is an HTTP header used for web browser cache validation. 

The historical data is stored in five directories named according to the type of data inside it. Each directory contains JSON files corresponding to the data described above.





\begin{figure*}[!htbp]
    \centering
    \includegraphics[width=0.25\textwidth]{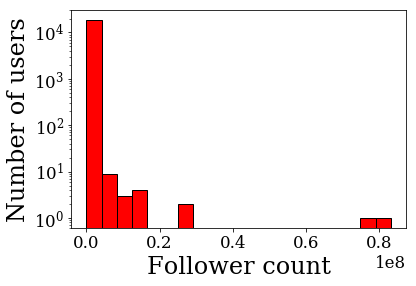}
    \hfill
    \includegraphics[width=0.24\textwidth]{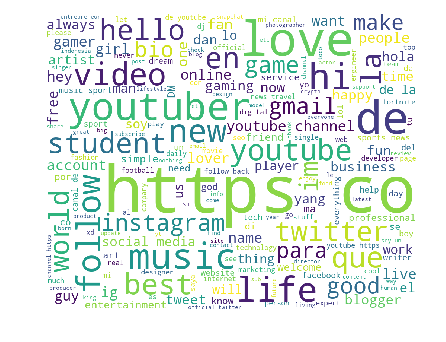}
    \hfill
    \includegraphics[width=0.25\textwidth]{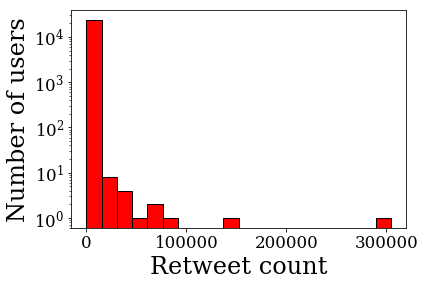}
    \hfill
    \includegraphics[width=0.24\textwidth]{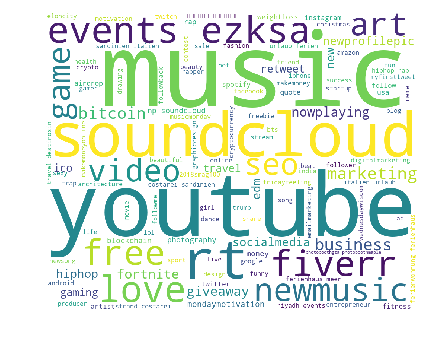}
  
    \caption{(a) Distribution of follower count, and (b) wordcloud aggregated over description text for users registered in blackmarket for collusive follower appraisals. (c) Distribution of retweet count, and (d) wordcloud aggregated over tweet text for tweets submitted in blackmarket for collusive retweet appraisals. Note that for clarity we remove common stopwords and single-letter words.}%
    \label{fig:distribution}
\end{figure*}

\subsection{Time-series Data}
We also collect the following time-series data for retweets and followers on Twitter:





\textbullet{} \texttt{user\_timeline}: This is a JSON list of tweet objects in the user's timeline, which consists of the tweets posted, retweeted and quoted by the user. The file created at each time interval contains the new tweets posted by the user during each time interval.

\textbullet{} \texttt{user\_followers}: This is a JSON file containing the user ids of all the followers of a user that were added or removed from the follower list during each time interval.

\textbullet{} \texttt{user\_followees}: This is a JSON file consisting of the user ids of all the users followed by a user, i.e., the followees of a user, that were added or removed from the followee list during each time interval.

\textbullet{} \texttt{tweet\_details}: This is a JSON object representing a given tweet, collected after every time interval.

\textbullet{} \texttt{tweet\_retweets}: This is a JSON list of tweet objects representing the most recent 100 retweets of a given tweet, collected after every time interval.

The time-series data is stored in directories named according to the timestamp of the collection time. Each directory contains sub-directories corresponding to the data described above.




\begin{table}[!htbp]
\centering
\begin{tabular}{ |c|c|c|c| } 
 \hline
 \textbf{User Type} & \textbf{Total} & \textbf{Suspended} &\textbf{Verified} \\ 
 \hline
 Retweet requests & 4,989 & 492 & - \\
 \hline
 Follower requests & 2,350 & 297 & 28 \\
 \hline
\end{tabular}
\caption{Summary of Twitter users for which time-series information was collected from Freemium blackmarket services.}
\label{table:timeseries_twitter}
\end{table}

\begin{figure*}[!htbp]
    \centering
    \includegraphics[width=0.32\textwidth]{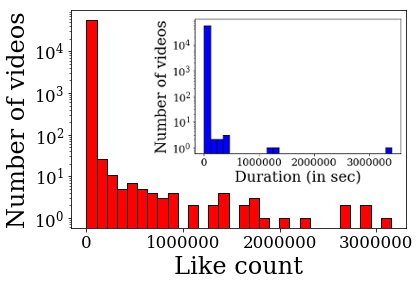}
    \hfill
    \includegraphics[width=0.32\textwidth]{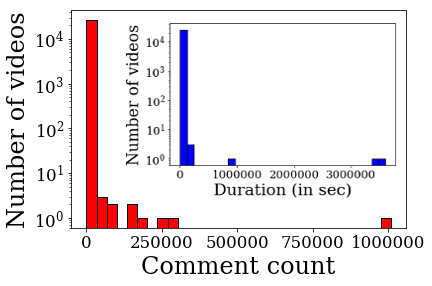}
    \hfill
    \includegraphics[width=0.32\textwidth]{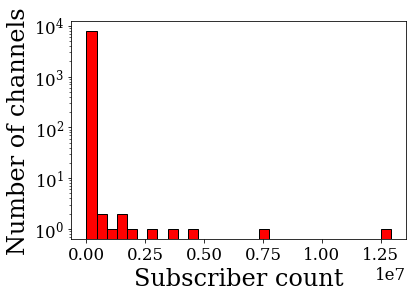}
    \hfill
    \caption{Distribution of (a) like count, (b) comment count, and (c) subscriber count for videos/channels submitted in blackmarket for collusive like, comment and subscription appraisals. The inset in (a) and (b) shows the distribution of the duration of the videos.}
    \label{fig:youtube_data}
\end{figure*}

Tables \ref{table:historical_twitter} and \ref{table:timeseries_twitter} detail the observations of the historical data and the time-series data for Twitter. It can be seen that only a very small fraction of the user accounts and tweets are no longer available on Twitter. We also observed some blackmarket customers who are marked as `Verified' by Twitter. Table \ref{table:freemium_youtube} details the observations of the historical data for YouTube collected from the blackmarket services. It can be seen that only a very small fraction of the channels and videos have already been removed from YouTube. This further motivates the need to develop techniques to analyze and detect collusive entities on online media platforms.






\section{Analysis of \dbname~Data}
In this section, we provide an analysis of  the \dbname~dataset to gain useful insights that will assist in demonstrating the opportunities opened by this new dataset.\\

\noindent\textbf{Twitter Data.} Figure \ref{fig:distribution}(a) and Figure \ref{fig:distribution}(b) shows the follower count distribution and wordcloud generated from the description text of users registered in blackmarket for collusive follower appraisals. We found that the maximum and average follower count was $83.28$ million and  $26432.28$ respectively. In Fig. \ref{fig:distribution}(b), we clearly see the presence of social media keywords such as ``follow", ``youtube", ``gmail" etc.  Figure \ref{fig:distribution}(c) and Figure \ref{fig:distribution}(d) shows the retweet count distribution and wordcloud generated from the text of tweets submitted in blackmarket for collusive retweet appraisals. We found that the maximum and average retweet count was $304442$ and  $131.08$ respectively. Also in Fig. \ref{fig:distribution}(d), along with the presence of social media keywords, we also observe some advertising keywords such as  ``free", ``seo" etc. For the collusive tweets submitted for collusive retweet appraisals, we analyze the machine-detected language of the tweet text using langdetect library\footnote{\url{https://pypi.org/project/langdetect/}}. We observe that $28.55\%$ of these tweets are written in non-english languages. The presence of multi-lingual tweets in the \dbname~dataset further adds to its contributions  that enable researchers to explore cross-lingual learning and also develop new tools for languages other than English for various NLP-based tasks in the area of anomaly detection research. \\

\begin{figure}[!t]
\centering
\includegraphics[width=0.8\columnwidth]{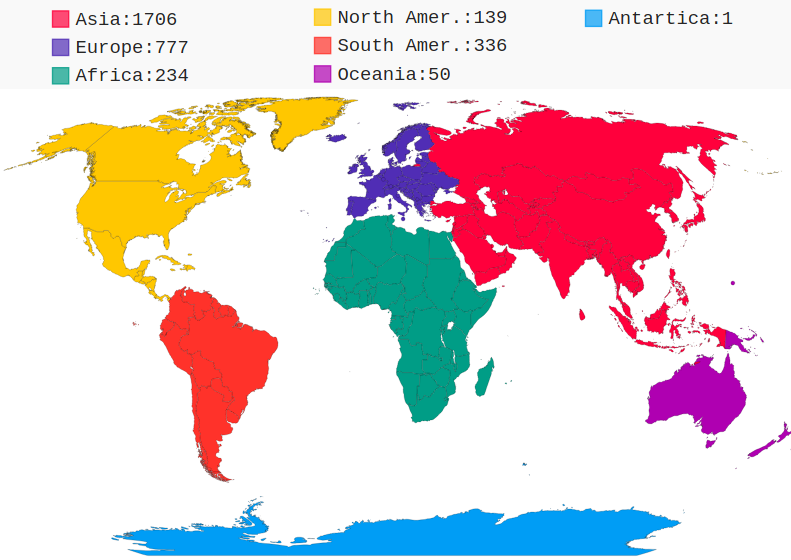}
\caption{Region-wise count of YouTube channels submitted to blackmarket services for collusive subscription requests. Asia is the top region accounting for 52.6\% of the total channels, followed by Europe, with 23.8\% of the total channels.}
\label{fig:youtube_region}
\end{figure}
\begin{figure}[!t]
    \centering
    \includegraphics[width=0.6\columnwidth]{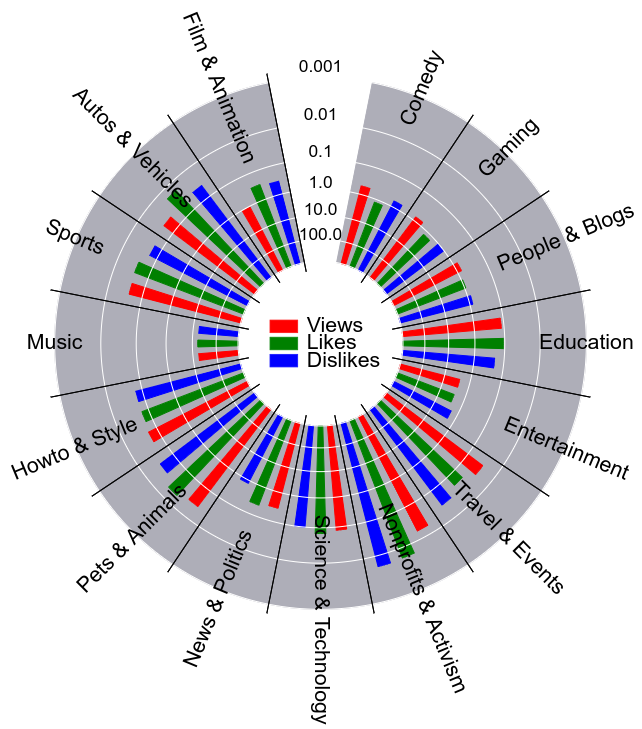}
    \caption{Genre-wise representation of views, likes and dislikes for YouTube videos registered in blackmarket services. `Non-profits \& Activism' is the top genre for collusive appraisals because of its three unique perks: (i) call to action overlays, (ii) a donation button, and (iii) Google Ad grants for paid advertising campaigns.}
    \label{fig:youtube_genre}
\end{figure}

\noindent\textbf{YouTube Data.} Figure \ref{fig:youtube_data}(a) show the distribution of (a) like count, (b) comment count, and (c) subscriber count for videos/channels submitted in blackmarket for collusive like, comment and subscription appraisals. The inset in (a) and (b) shows the corresponding distribution of the duration of the videos. Figure \ref{fig:youtube_region} shows the number of channels submitted to blackmarket services for collusive subscription requests from different regions. We first extract the \texttt{country} parameter for each YouTube channel in our dataset and convert into a new parameter \texttt{continent} using the PyCountry library\footnote{\url{https://pypi.org/project/pycountry/}}. Asia is the top region accounting for 52.6\% of the total channels, followed by Europe, with 23.8\% of the total channels. For the videos submitted for collusive comment requests, we measured the sentiment of the comments received by the videos. The sentiment was measured using Python TextBlob library\footnote{\url{https://textblob.readthedocs.io/en/dev/quickstart.html}}. Unsurprisingly, we found that more than $95\%$ of the comments received by these videos have a positive sentiment. 
Figure \ref{fig:youtube_genre} shows a genre-wise representation of the count of views, likes, and dislikes for YouTube videos. Most of the videos for collusive requests are from the genre `Non-profits \& Activism.' The possible reason behind such a trend is that this genre allows organizations to upload videos with free premium services such as donate buttons, call-to-action overlays, live-streaming, and goal tracking, which are preferred ways to reach new as well as old audiences.  \\



\section{SearchBM: A search engine for collusive entity discovery}
To better aid the collusive entity discovery process and provide a better understanding of how and where the entities have been used, we developed \texttt{SearchBM}, an API and a web portal for our end-users to effectively query within the \dbname~dataset. Users can directly search for a query text using the interface of \texttt{SearchBM}. The query is then sent to our backend server. The backend of the server is developed using Python-Flask (\url{http://flask.pocoo.org/}). Currently, the API can accept a given text at the following request paths: \\


\noindent \textbullet{} {(\texttt{<text>,/collusive\_twitter\_retweets})}: This checks for presence of the query text in the tweets submitted in blackmarket services for gaining collusive retweets.

\noindent \textbullet{} {(\texttt{<text>,/collusive\_yt\_channels})}: This checks for presence of the query text in the description of YouTube channels submitted in blackmarket services for gaining collusive subscriptions.

\noindent \textbullet{} {(\texttt{<text>,/collusive\_yt\_likes})}: This checks for presence of the query text in video description of YouTube videos submitted in blackmarket services for gaining collusive likes.

\noindent \textbullet{} {(\texttt{<text>,/collusive\_yt\_comments})}:
This checks for presence of the query text in the video description and user comments of YouTube videos submitted in blackmarket services for gaining collusive comments.\\ \\

\begin{figure}[!htbp]
    \centering
    \includegraphics[width=0.4\columnwidth]{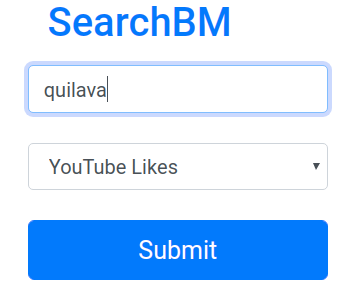}
    \hfill
    \includegraphics[width=0.4\columnwidth]{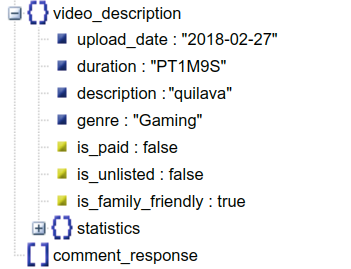}
    \hfill
    \caption{\texttt{SearchBM} entity query form. End-users have to enter the query text in the textbox and select one of the types from the drop-down menu.}
    \label{fig:search_bm}
\end{figure}

The API returns a JSON object indicating the presence of the text in our datasets. If the entity is found in our database, the API returns the details of the entity from our dataset. Note that to maintain anonymity, we only show specific attributes of the entity and not the identifiers (tweet/user identifier for Twitter data and video/channel identifier for YouTube). We also provide a web interface where end-users can enter the query text and select one of the services - \textit{Twitter retweets, YouTube likes, YouTube comments and YouTube subscriptions} and get the corresponding details via the API. Fig. \ref{fig:search_bm}(a) shows the interface of \texttt{SearchBM}. End-users have to enter the query text in the textbox and select one of the types from the drop-down menu. On clicking the submit button, the query parameters are sent as a GET request to our API, which returns a JSON object indicating the presence of text in the collusive entity. The front-end uses a JSON viewer, as shown in Fig. \ref{fig:search_bm}(b) to display the entity details.
\section{Research Opportunities using the \dbname~dataset}
We believe that \dbname~can can benefit in many threads of anomaly detection research. We discuss a few examples below:\\

\begin{enumerate}
\item \textbf{Fraudulent user/entity detection:} A great deal of work has been devoted to fraudulent user/entity detection in online media platforms. The task of detecting fraudulent users includes identifying fake users \cite{gupta2013faking,atodiresei2018identifying,fire2014friend}, spammers  \cite{miller2014twitter,benevenuto2010detecting}, bots \cite{chavoshi2016debot,chavoshi2016identifying,dickerson2014using}, collusive users \cite{dutta2018retweet,dutta2019blackmarket,chetan2019corerank,arora2019multitask,dutta2020hawkeseye}, sockpuppets \cite{kumar2017army} etc. Most of the above algorithms only detect individual users. However, in reality, it is seen that the anomalous phenomena also occur in groups. The group detection task (finding a group of users that jointly exhibit fraudulent behavior) is more difficult as compared to the individual detection task due to the variation present in the inter-group dynamics. In our case, the users of freemium blackmarket services perform actions (retweet/like/comment) on collusive entities in order to gain credits. Therefore, it is almost certain that users in \dbname~must have interacted with each other in order to gain credits. We believe that the availability of the \dbname~dataset can foster fraudulent user/entity detection approaches (both individual and group) with the advantage of adding the topical as well as the temporal dimension. 

\item \textbf{Mining connectivity patterns:} Understanding connectivity patterns of an underlying network is a well-studied problem in the literature.  It include tasks such as inferring lockstep behavior \cite{beutel2013copycatch}, dense block detection \cite{shin2016m}, detecting core users  \cite{shin2016corescope,shin2018patterns}, identifying the most relevant actors in a network  \cite{borgatti2006identifying}, sudden  appearance/disappearance of links \cite{eswaran2018spotlight} etc. Using the \dbname~dataset, researchers can create various networks among the users/entities present in the dataset to investigate various structural patterns of the network.

\item \textbf{Modeling temporal evolution:} As \dbname~dataset contains time-series data, it can be used for various temporal modeling tasks. Some example tasks include detecting time periods containing unusual activity \cite{giatsoglou2015retweeting}, identifying repetitive patterns in time-evolving graphs \cite{zeidanloo2010botnet} etc.

\item \textbf{Diffusion modeling:} The \dbname~dataset can be represented as networks based on the actions performed by the collusive users on the content of other users of the services. It could be then use to study multiple diffusion modeling tasks such as influence maximization (selecting a seed set to maximize the influence spread) \cite{jendoubi2017two,mei2017influence}, predicting information cascade \cite{rattanaritnont2011study}, measuring message propagation and social influence \cite{ye2010measuring,brown2011measuring} etc.

\item \textbf{Event-specific studies:} As \dbname~contains data obtained from multiple sources and spans over a long period of time, it may consist of information from many major events \cite{atefeh2015survey}, which can be easily extracted for event-centric studies. Researchers can also check how these users/entities were involved in manipulating the popularity of events by artificially inflating the social growth of users/entities in online media \cite{zhang2016twitter}.

\item \textbf{Multi-lingual studies:} In the previous section, we mentioned the presence of multilingual texts in the \dbname\ dataset. We anticipate that the multilingual data will be useful for a broad range of Natural Language Processing (NLP) tasks in the anomaly detection domain.
\end{enumerate}

\section{How \dbname~is a FAIR-compliant dataset?}
In this section, we explain how we have made the \dbname~dataset compliant to the four FAIR data principles: \textit{Findable, Accessible, Interoperable and Re-usable}.

To make the \dbname~dataset \textit{Findable} and \textit{Accessible}, our dataset is publicly available on Zenodo which allows downloading the entire dataset with the following citation: Hridoy Sankar Dutta, Udit Arora \& Tanmoy Chakraborty. (2021). ABOME: A Multi-platform Data Repository of Artificially Boosted Online Media Entities [Data set]. Zenodo. \url{http://doi.org/10.5281/zenodo.4437987}.

To make the \dbname~dataset \textit{Interoperable} and \textit{Re-usable}, the dataset files are provided in standard JSON (JavaScript Object Notation) format that can be easily parsed using any standard JSON parser and can be exported to other data formats like CSV (Comma Separated Values), XML (Extensible Markup Language) etc. We also provide a readme file to optimize the re-use of the dataset.

\section{Conclusion}
Collusive entity detection is an important problem that  has largely been overlooked. To the best of our knowledge, \dbname~is the first dataset in the literature that consists of multi-platform data related to blackmarket-driven collusive entities collected from two credit-based freemium services - YouLikeHits and Like4Like. In addition, we also designed an API and a web portal, \texttt{SearchBM} to discover collusive entities using text search queries. We believe that the datasets released in this paper will provide more opportunities for researchers to advance the development of technologies in detecting collusive entities in online media, thereby creating an adequate social space. We also encourage researchers working in the domain of privacy and security in OSNs to propose interesting tasks and use \dbname\ as a benchmark.

\bibliographystyle{unsrt}
\bibliography{references}
\end{document}